\documentclass[aps,prl,floatfix,twocolumn,superscriptaddress,showpacs]{revtex4}

\pdfoutput=1
\usepackage{graphicx}
\usepackage{amssymb}
\usepackage{amsmath}
\usepackage{color}

\newcommand{\bea}{\begin{eqnarray}}
\newcommand{\eea}{\end{eqnarray}}
\newcommand{\be}{\begin{equation}}
\newcommand{\ee}{\end{equation}}
\begin{document}
\title{Dynamical disentangling and cooling of atoms in bilayer optical lattices}

\author{A.~Kantian}
\affiliation{Nordita, KTH Royal Institute of Technology and Stockholm University, Roslagstullsbacken 23, SE-106 91 Stockholm Sweden}
\author{S.~Langer} 
 \affiliation{Department of Physics and Astronomy, University of Pittsburgh, Pittsburgh, Pennsylvania 15260, USA}
\author{A.~J.~Daley} 
\affiliation{Department of Physics and SUPA, University of Strathclyde, Glasgow G4 0NG, UK}
\date{\today}


\begin{abstract}
We show how experimentally available bilayer lattice systems can be used to prepare quantum many-body states with exceptionally low entropy in one layer, by dynamically disentangling the two layers. This disentangling operation moves one layer - subsystem $A$ - into a regime where excitations in $A$ develop a single-particle gap. As a result, this operation maps directly to cooling for subsystem $A$, with entropy being shuttled to the other layer. For both bosonic and fermionic atoms, we study the dynamics of this process, and show that disentangling can be realised cleanly in ongoing optical lattice experiments. The corresponding entanglement entropies are directly measurable with quantum gas microscopes, and as a tool for producing lower-entropy states, this technique opens a range of applications beginning with simplifying production of anti-ferromagnetically ordered states of fermions. 
\end{abstract}
\maketitle

Understanding entanglement in many-body systems \cite{Amico2008,Eisert2010} provides a new way to characterise a variety of phenomena, ranging from the identification of topological states \cite{Jiang2012,Isakov2011,Kitaev2006,Levin2006} to out-of-equilibrium quench dynamics  and fundamental issues such as thermalisation \cite{Kaufman2016,Rigol2008}. Measures of entanglement in many-body systems can also be directly accessed in experiments, as was recently demonstrated for R\'enyi entropies of itinerant atoms in an optical lattice \cite{Islam2015,Daley2012,Moura-Alves2004}. In the present work, we show how dynamical manipulation of entanglement for atoms in bilayer optical lattices could be used as a tool to address a key experimental challenge. Based on processes that result in a dynamical disentangling of two layers within a bilayer optical lattice at low temperatures, as shown in Fig.~1, it is possible to realise regimes where most of the thermal entropy in the system is transferred into one of the two layers. This produces one low-entropy layer which can be further adiabatically manipulated to access a broad range of low-temperature phenomena that are presently unachievable. Moreover, all operations required for such a dynamical disentangling are readily available in experiments with optical superlattices \cite{Folling2007} or quantum gas microscopes \cite{Preiss2015}.

The first milestone in this direction would be the simplified preparation of quantum magnetic ordering driven by super-exchange processes, which is challenging due to the small energy gaps involved \cite{Capogrosso-Sansone2010,Ho2007,Jordens2010}. Recent seminal experiments detecting short-range anti-ferromagnetic correlations for fermionic atoms in optical lattices \cite{Greif2013,Hart2015} demonstrated entropies within a factor of two of that required for the N\'eel transition, and further progress has been made with individual site addressing in quantum gas microscopes \cite{Boll2016,Parsons2016}, revealing magnetic correlations in 2D on length scales up to eight lattice sites. However, with the eventual goals of observing effects that require much lower temperatures still \cite{Bloch2012,Cirac2012}, it is imperative to develop new ways to strongly reduce the entropy. We show below that our scheme could reduce entropies by around an order of magnitude starting from initial states attainable in current experiments for bosons, and can be combined with dimerised lattices to produce low-entropy states with magnetic ordering for Fermions.


Below we first provide an intuitive explanation of dynamical disentangling by considering bosons in an optical lattice. We then look at the specific applications to producing magnetically ordered states of fermions, before discussing the generalisation of this idea to other systems.

\begin{figure}[t]\label{fig:Fig1}
\includegraphics[width=0.5\textwidth]{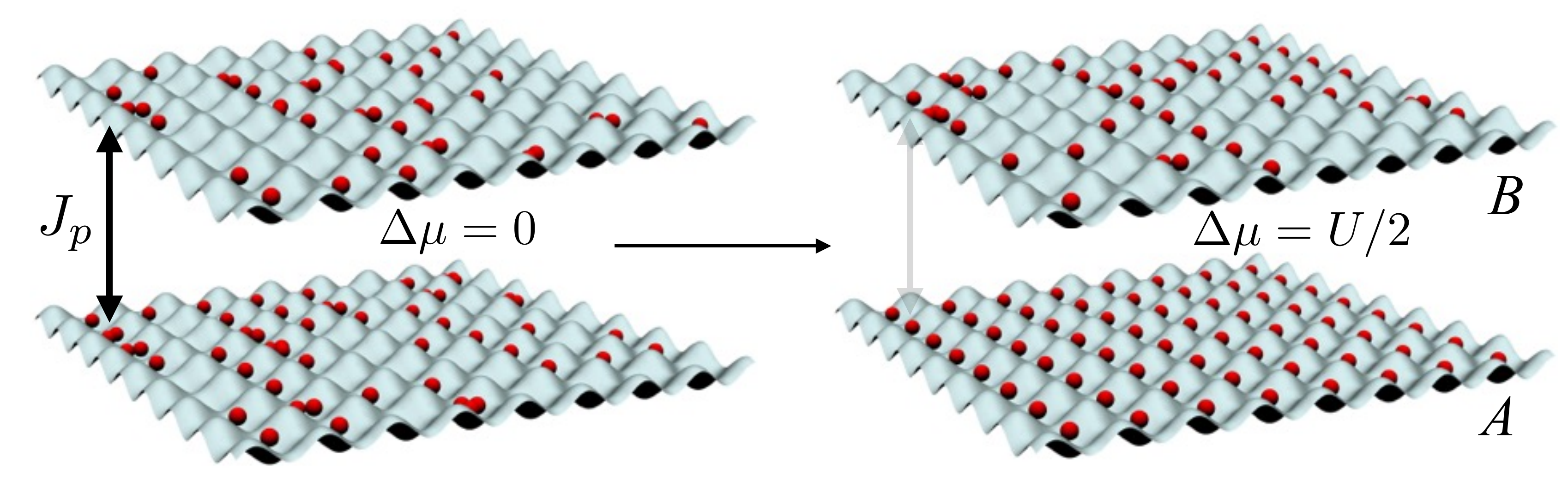}
\caption{(Color online) Basic schematic of dynamical disentangling in a bilayer scheme for single-component bosons. (a) Two tunnel-coupled layers (with interlayer tunnelling $J_p$) are prepared at the same chemical potential (identical trap depths in the vertical direction). In this regime, particles are delocalised between the layers, which are entangled at zero temperature. By increasing interactions, manipulating the relative chemical potential of the layers, and then removing the tunnel coupling, layer $A$ can be prepared in a Mott-insulating state. Having a single gapped state in layer $A$ strongly suppresses entanglement of the two layers at zero temperature. At non-zero temperatures, entropy per particle is much higher in layer $B$, where atoms are free to move.}
\label{fig:setup}
\end{figure}

\emph{Intuitive picture of dynamical disentangling for bosons in a bilayer system --}
Here we introduce the concept of dynamical disentangling of two subsystems by considering bosons in a bilayer optical lattice, as depicted schematically in Fig.~1. For atoms in the lowest Bloch bands in each layer, under well-controlled approximations the hamiltonian is a Bose-Hubbard model \cite{Lewenstein2012,Jaksch2005,Jaksch1998}, $H=H_A+H_B+H_c$, where ($\hbar\equiv 1$) 
\begin{eqnarray}
H_{X}= -J \sum_{\langle{l,k}\rangle} b^\dag_{l,X}b_{k,X} + \frac{U}{2}\sum_{l}n_{l,X}(n_{l,X}-1),\\
H_{c}(t)=\sum_l\left[-J_p(t)(b^\dag_{l,A}b_{l,B}+b^\dag_{l,A}b_{l,B})-\Delta \mu(t) n_{l,A}\right].
\end{eqnarray}
Here, $X\in\{A,B\}$, $b^\dag_{l,X}$ creates a boson on site $l$ in layer $X$, $J_p(t)$ denotes the hopping from one layer to the other and $\Delta\mu(t)$ is a global energy shift between the layers. Within each layer, $n_{l,X}=b^\dag_{l,X} b_{l,X}$, the tunnelling amplitude is $J$, and the onsite interaction shift is given by $U$. This system can be realised, e.g., either in a quantum gas microscope \cite{Preiss2015} or by using superlattices \cite{Folling2007}.

If we choose the number of particles $N$ to be fewer than the sum of lattice sites of both layers, $M\equiv M_A+M_B$, then for $\Delta \mu=0$ the zero-temperature ground state will involve the atoms being delocalised between the two layers. This results in entanglement of the two subsystems corresponding to layers $A$ and $B$. Thus, even though the total system is in a pure state with entropy $S\equiv -{\rm Tr}\{\rho \log \rho\}=0$, the entropy of the reduced subsystem for layer $A$, $S_A\equiv -{\rm Tr}\{\rho_A \log \rho_A\}$, will be non-zero, $S_A>0$, where $\rho$ is the density matrix for the whole system and $\rho_A={\rm Tr}_B\{\rho\}$~\cite{Islam2015,Dowling2006}. We now consider what happens for weak coupling between the layers, $J_p \rightarrow 0$. If we increase the difference in chemical potential between the layers, $\Delta \mu$, then we can favour the transfer of particles to layer $A$. As depicted schematically in Fig.~1, and in the mean-field phase diagram in Fig.~2a, for sufficiently large $U/J$, we can enter a regime where one layer will remain in a superfluid (SF) regime at zero temperature, but the other will enter the gapped Mott Insulator (MI) regime. At zero temperature, the gap will suppress excitations in layer $A$, and for $J_p\rightarrow 0$, that layer will be in its ground state. Contributions from other states in layer $A$ will be suppressed by the excitation gap, leading to a suppression of entanglement $S_A$, because mostly just a single state of subsystem $A$ contributes to the ground state of the whole system.

\begin{figure}[t]
\includegraphics[width=0.5\textwidth]{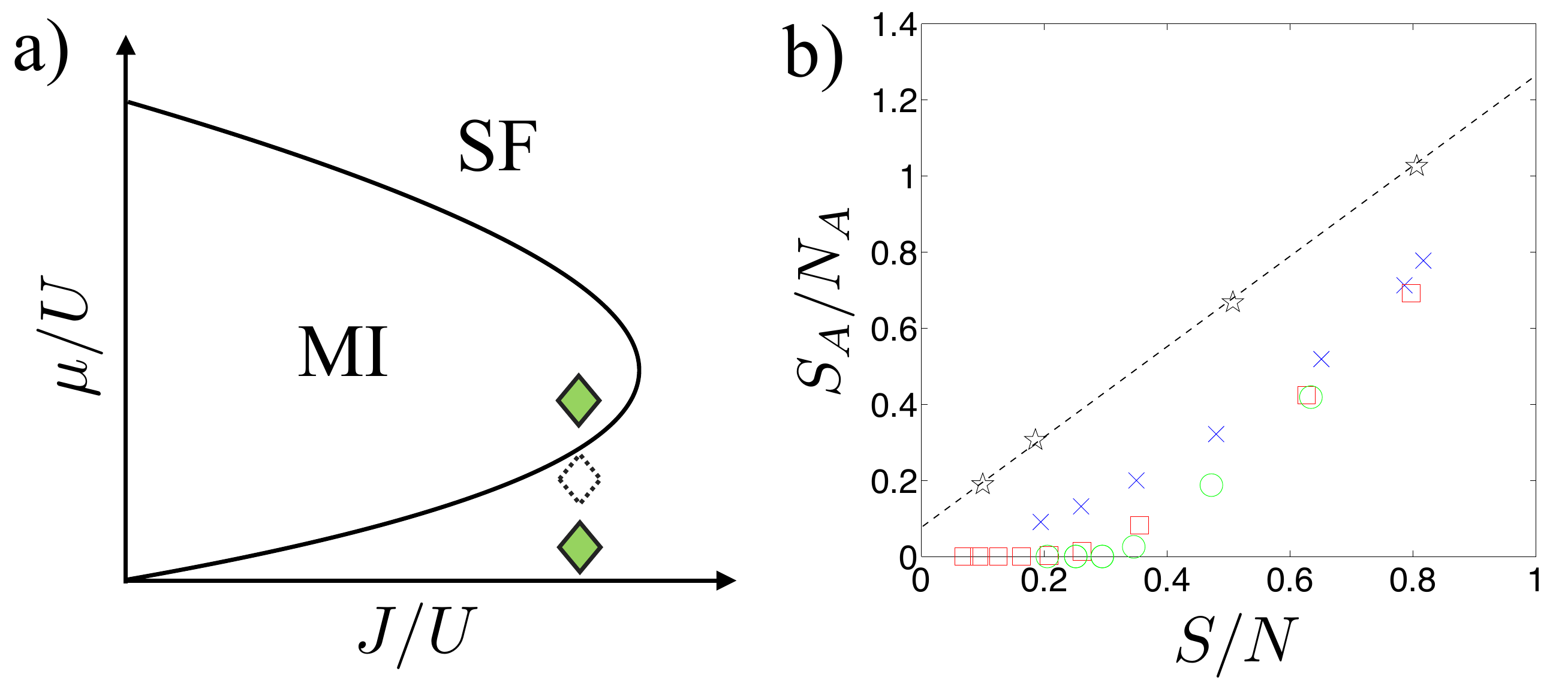}
\caption{(Color online). Equilibrium description of the disentangling scheme. (a) The changing chemical potentials in the process from Fig.~1 seen on the zero-temperature mean-field phase diagram for the Bose-Hubbard model in the local density approximation. Two superfluid layers with less than unit filling (dotted diamond) are separated in chemical potential so that one becomes Mott Insulating and the other remains superfluid (solid diamonds). The separation between the chemical potentials $\mu$ of the two layers corresponds to $\Delta \mu$. (b) Entropy per particle in layer $A$ of a 1D bilayer system of length $6$ with $8$ lattice bosons with onsite repulsion $U$,
as a function of total entropy per particle. The black stars, with the fitted dotted line, show the entropy per particle in one layer when $U/J=5$ and the potential offset $\Delta\mu=0$. The intercept at $S/N=0$ highlights the non-zero entanglement entropy of the two layers for zero temperature. The remaining points show target parameters for the dynamical disentangling operation, with $\Delta\mu=U/2$. Blue crosses denote $U/J=8$, red squares $U/J=20$, and green circles $U/J=50$. For large $U/J$, the entropy per particle in layer $A$ is strongly suppressed as the total entropy per particle is lowered. }
\label{fig:Fig2}
\end{figure}

To evaluate the impact of higher temperatures on this effect, we calculate via exact diagonalisation the subsystem entropy per particle $S_A/N_A$ ($N_X\equiv\sum_l\langle n_{l,X}\rangle$) and entropy per particle of the whole system $S/N$ at different temperatures, and plot these per-particle entropies against each other for a 1D system with 8 particles in 12 lattice sites ($M_{A,B}=6$) in Fig.~\ref{fig:Fig2}b. When $S/N$ is large, layer $A$ is indeed measurably entangled with layer $B$, but as $S/N$ is reduced, the entropy is almost entirely transferred to layer $B$, as $S_A$ is exponentially suppressed. At zero temperature, $S\rightarrow 0$, we see directly the suppression of entanglement between the layers by comparing the black stars, which show $S_A$ for $\Delta \mu=0$ with the other curves, where $\Delta\mu=U/2$, and $S_A\rightarrow 0$ as $S\rightarrow 0$. We then see how entropy can, in principle be dynamically transferred from layer $A$ to layer $B$ in the process depicted in Fig.~1, as in the ideal case, the initial entropy associated with a black stars in layer $A$ can be dynamically reduced up to the values indicated by the other markers. As a means to reduce the entropy in one region of a trapped atomic gas, this is reminiscent of the movement of entropy from a central region in a purposefully-designed trapping potential, as proposed in Ref.~\cite{Bernier2009}. However, The bilayer geometry of the present scheme makes it much more straight-forward to isolate the high- and low-entropy subsystems from one another, and should yield improved timescales, given that mass transport in dynamical disentangling only needs to be local.  

This intuitive picture for why a disentangling operation should yield low-entropy states in layer $A$ when that develops a single-particle gap, which we considered up to here in the static limit and for $J_p \rightarrow 0$, requires a non-trivial justification as soon as $J_p \neq 0$. This coupling could conceivably result in long-range correlations in layer $A$ through layer $B$, such that we no longer have a decoupled MI. It is thus of central importance to know whether the coupling along the boundary between the layers can involve exponentially many states at non-vanishing weight, which would result in large entanglement. However, this can be shown not to be the case~\cite{SM}. Namely, entanglement is small, with the number of states participating in it scaling \textit{linearly} and \textit{not} exponentially in $M_A$, and scaling to zero with $J_p/\delta$ whenever $H_c$ is local and generates only single-particle excitations in layer $A$, which is the case of the MI states we discussed above. More complex still are questions concerning the dynamics: as the whole system is initially ungapped (and layer $B$ is always ungapped), we need to check if the dynamical ramps can still produce low-entropy states in layer $A$. In the following, we treat examples of the dynamics that show for finite systems it is possible to perform these ramps adiabatically at zero temperature, and that at non-zero temperature, the vast majority of the entropy is still transferred to layer $B$ even if the ramp is not adiabatic. 

\emph{Time-dependence of dynamical disentangling for bosons --} We first investigate the adiabaticity of a ramp beginning with particles delocalised over two layers into the disentangled state at zero temperature. In Fig.~3a, we plot the final many-body-state fidelities when we consider two coupled 1D chains, where we can compute the dynamics using adaptive time-dependent density matrix renormalization group techniques \cite{Schollwoeck2011,Vidal2004,White2004,Daley2004,Verstraete2008}. We see that relatively short ramps, with a timescale $T\approx 20J^{-1}$, the fidelity $F(T)=|\langle \psi(T)|\psi_{\rm target} \rangle |^2$ of the final state of the ramp with $J_{p}\rightarrow 0$ and $\Delta\mu=U/2$, $\psi_{\rm target}$ to the time-evolved state $|\psi(T)\rangle$ is almost one. For zero temperature, such high fidelities are a consequence of finite-size gaps, and the required timescale increases as the system size grows. 

At non-zero temperatures, we expect that the ramp will never be entirely adiabatic. However, if it is sufficiently slow we nonetheless expect that an increase in excitations primarily affects the final state in layer $B$, where the excitations are ungapped, with the gapped state in layer $A$ still protected. This can be particularly enhanced if we ensure optimal conditions for thermalisation between the layers during the ramp. To demonstrate this, we show in Fig.~3b the final per-particle entropy of layer $A$, $(S_A/N_A)_{final}$ as a function of the initial per-particle entropy $(S/N)_{initial}$ of the whole system for a small system that still permits propagation of density matrices via exact diagonalisation. We note that $(S_A/N_A)_{final}/(S/N)_{initial}$ is strongly suppressed, and that even with the moderate ramp times $\sim 100 J^{-1}$, it is possible to achieve a lowering of $(S_A/N_A)_{final}$ by an order of magnitude over the initial $(S/N)_{initial}$. Because we expect some degree of non-adiabaticity, the final entropy depends in general on the choice of ramp. We compare two ramps, one with $U/J=20$ fixed throughout the ramp, and one in which we initially have a small value of $U/J$. The latter case promotes thermalisation between the layers, and results in a substantially lower value for $(S_A/N_A)_{final}$.

\begin{figure}[t]
\begin{center}
\includegraphics[width=0.5\textwidth]{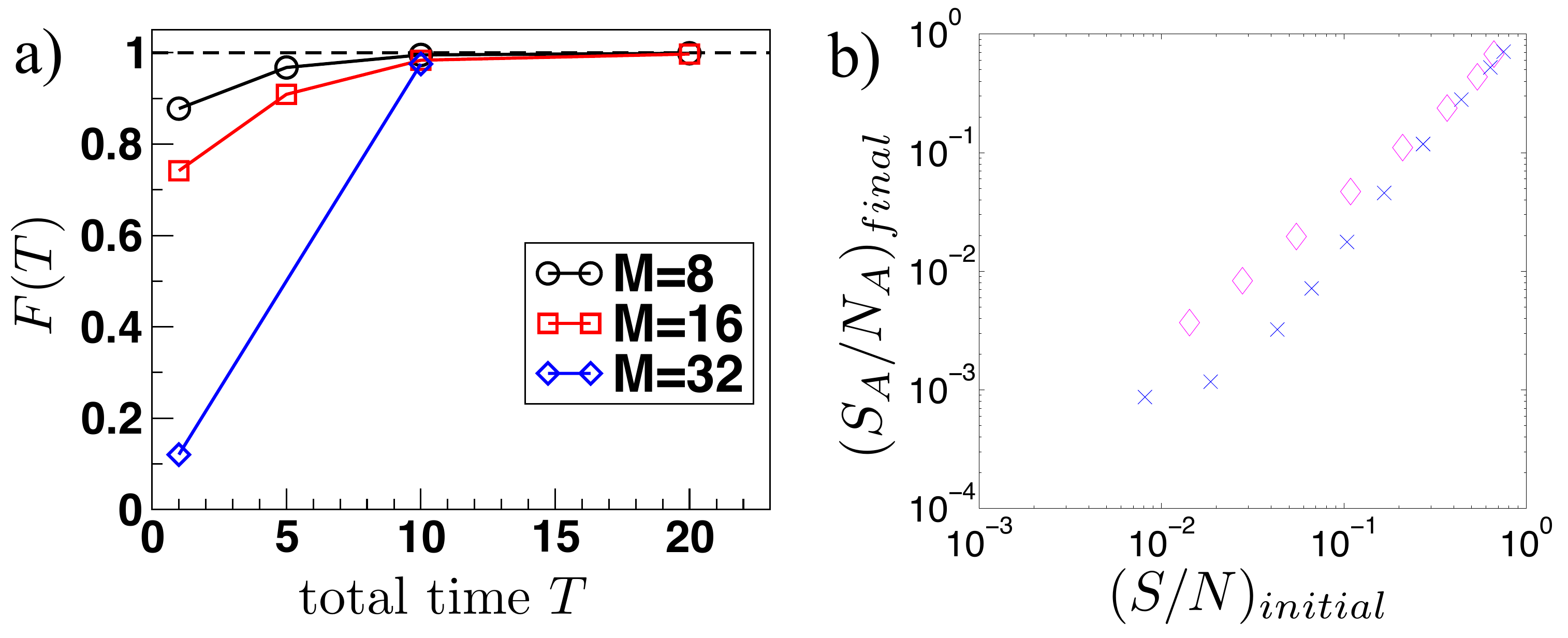}
\caption{(Color online) Analysis of the time-dependent disentangling operation for bosonic atoms. (a) For zero temperature, we show the fidelity of the final state of the ramp to the ground state of the system with $\Delta\mu=U/2$, and then to $J_p=0$, against total ramp time $T$ (measured in units of $J^{-1}$), computed using t-DMRG techniques for up to $M=16\times 2$ lattice sites, always taking $N=3M/4$ bosons at $U=8J$. We ramp parameters in two stages, beginning in the ground state with $\Delta\mu=0$ and $J_p=J$, first ramping linearly in time to $\Delta\mu=U/2$, and then to $J_p=0$. (b) Final entropies per particle in the layer $A$ with non-zero temperatures in a 1D bilayer system with $M=5\times 2$, $N=7$. We compare the final entropies per particle in layer $A$ at the end of two possible disentangling ramps against total initial entropy per particle, $S$. Here, $\Delta\mu$ is linearly increased from $0$ to $10J$ over a time $92J^{-1}$, and then subsequently $J_\perp$ is linearly lowered from $=J$ to $=0$ within time $4J^{-1}$. The blue crosses show a ramp with $U$ initially kept at a low constant value $U=J$ for a time $30J^{-1}$, and is subsequently linearly ramped up to $U=20J$ within a timespan of $62J^{-1}$, while the magenta diamonds show the same protocol, but with $U=20J$ constant as the other parameters are varied.}
\label{fig:boson_state}
\end{center}
\end{figure}

\emph{Dynamical disentangling and realisation of magnetically-ordered states for fermions --}
This scheme can be readily generalised to other states where we can induce a gap for excitations in one subsystem, including magnetically ordered states of multi-component bosons or fermions in optical lattices. This would be a crucial step in simplifying the production of magnetically ordered states, e.g., antiferromagnets in the Hubbard model at half filling. Again, we consider two layers, and for fermions with two spin states, the hamiltonian $\tilde H=\tilde H_A+\tilde H_B+\tilde H_c$, where ($\hbar\equiv 1$) 
\begin{equation*}
\tilde H_{X}= - \sum_{\langle l,k \rangle ,\sigma } J_{lk} c^\dag_{l\sigma ,X}c_{k\sigma ,X} + U\sum_{l}n_{l\uparrow ,X}n_{l\downarrow ,X},
\end{equation*}
\begin{equation*}
\tilde H_{c}(t)=\sum_{l,\sigma} \left[-J_p(t)(c^\dag_{l\sigma ,A}c_{l\sigma,B}+H.c.)-\Delta \mu(t) n_{l\sigma ,A}\right].
\end{equation*}
Here, $X\in\{A,B\}$, $c^\dag_{l\sigma ,X}$ creates a fermion of spin $\sigma$ on site $l$ in layer $X$, and $n_{l\sigma ,X}=c^\dag_{l\sigma ,X} c_{l\sigma ,X}$.
In principle, it is possible to apply the protocol we previously used for the bosons, taking two uniform layers - i.e. with the  tunnelling amplitude $J_{lk}$ being constant. However, because magnetic ordering is induced by a superexchange gap $\sim J^2/U$ the separation of timescales between $J$ and this gap can make our numerical calculations very challenging. 

Instead, we consider the dimerised lattice geometry that was recently realised by Greif et al., \cite{Greif2013}, and is depicted for a 1D case in Fig.~4a. In Ref.~\cite{Greif2013}, the equilibrium entropy in the presence of a varying trapping potential was studied with a strong coupling expansion, showing a potential redistribution of entropy to areas of lower chemical potential, as we see with the Bose-Hubbard model. If we take a bilayer system in this geometry, this will lead to strong suppression of the entropy per particle in the layer with lower chemical potential. 
In Fig.~4b, we demonstrate the adiabaticity of a chemical potential ramp in such a dimerised lattice, in analogy with Fig.~3a for bosons, beginning with less than half filling for the whole system, and producing a half-filled layer with spin singlets in each dimer. To characterise this final state, we plot the strength of the local dimer correlation functions as a measure of the final state. For all tested system sizes $L=8$ (blue circles), $L=16$ (red squares) and $L=32$ (green diamonds) this scheme exhibits clear power-law scaling to such low values that represent near-perfect spin-singlets prepared on each pair of sites in layer $A$. Based on the results of strong-coupling expansions in Ref.~\cite{Greif2013} and ED calculations, we see that the potential reduction in entropy is similar to that seen for Bosons in Fig.~2b. At current experimental entropies this would allow reductions of the order of a factor of two for easier entry into magnetically ordered states, with much larger reductions possible for lower entropy starting points.


As indicated in Fig.~4c, we then consider a low-entropy layer, such as we produced in layer $A$ above, as a starting point for realising a state with long-range anti-ferromagnetic order by increasing the coupling between dimers time-dependently, in analogy with Ref.~\cite{Lubasch2011}. 
Initially one has prepared one up- and one down-spin fermion with $U/J\gg 1$ on each pair of sites with tunnelling amplitude $J$ between them in their ground-state (i.e. the unique singlet state), while inter-dimer tunnelling $J_{ID}$ is at or near zero. Ramping $J_{1D}$ up to $J$ near-adiabatically should result in a smooth crossing over to the desired globally antiferromagnetic ground state of the Hubbard model at half-filling, as one is initially protected against coupling to excited states by the finite spin-gap. In Fig.~4d we demonstrate this, plotting one minus the fidelity as a function of the total ramp time after an exponential ramp, as detailed in the figure caption. The conclusion is that the low-entropy dimer state that can be achieved by dynamical disentangling can then be used to prepare a long-range antiferromagnet.

\begin{figure}[t]
\begin{center}
\includegraphics[width=8cm]{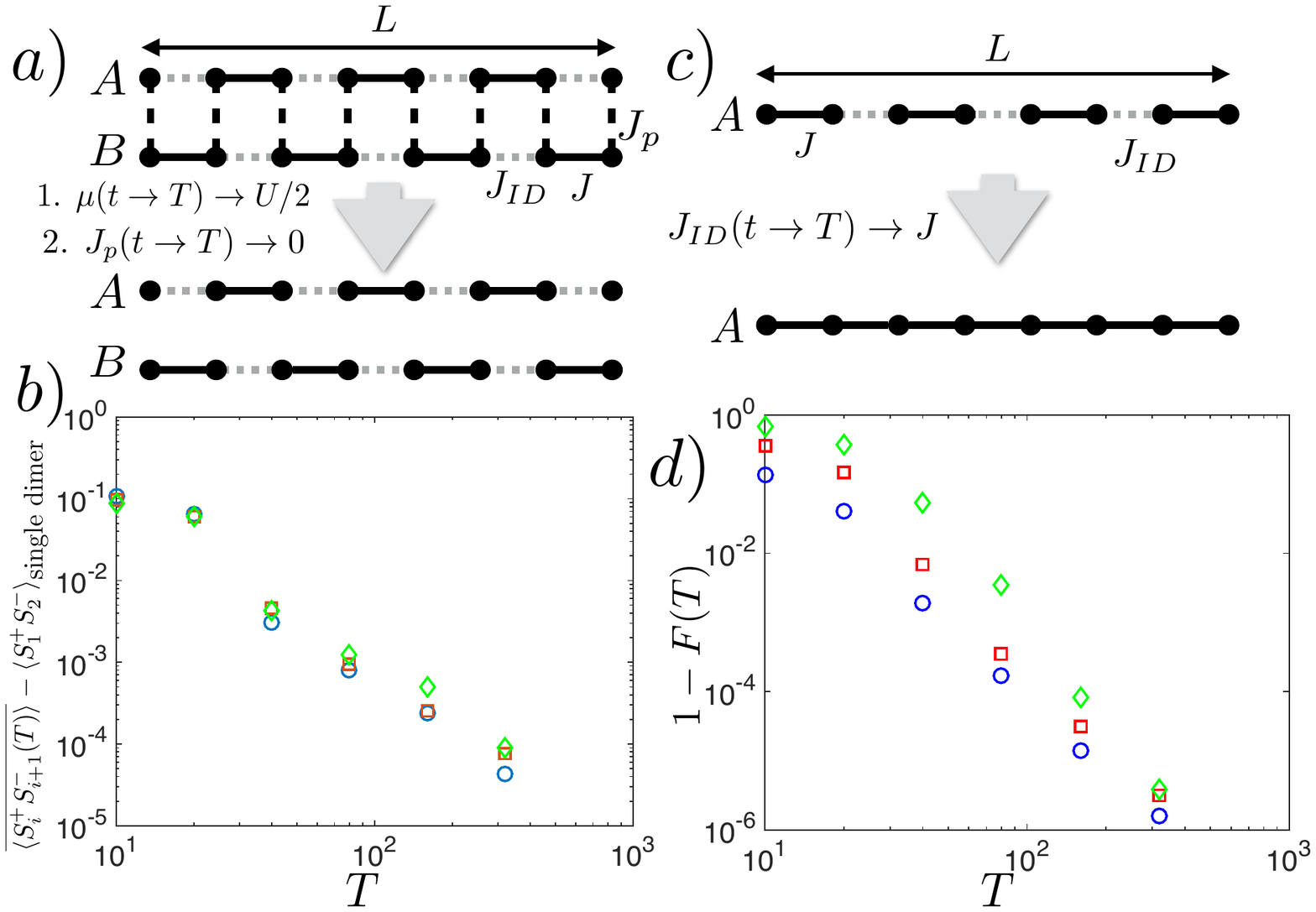}
\end{center}
 \caption{(Color online) (a) Overview of bilayer disentangling scheme for fermions: layers $A$ and $B$ are connected with tunnelling $J_p$. In each layer there are pairs of sites with tunnelling amplitude $J$ between them, which are connected with each other by inter-dimer tunnelling amplitude $J_{ID}$. (b) Result of ramping scheme shown in (a) for initial parameters $J_p=J$ and $J_{ID}=0$, where a potential difference between layers $A$ and $B$ is linearly ramped up to value $\Delta\mu=U/2$ in time $T$, then $\Delta\mu$ is kept fixed and  $J_p$ is linearly ramped to zero in time $T$. We show the difference of the average correlation over all dimers in $A$ $\overline{\langle S^+_{i}S^-_{i+1}\rangle}$ at the end of the ramp to the value on a single dimer with one spin-up and spin-down fermion, for system sizes $L=8$ (blue circles), $L=16$ (red squares) and $L=32$ (green diamonds). Here, $U/J=8$ and $N_\uparrow=N_\downarrow=3L/4$. (c) Schematic overview of antiferromagnetic state preparation starting from the final state of Fig. 4b, i.e. isolated pairs of singlets at zero temperature, by increasing tunnelling between dimers adiabatically. (d) Plot of $1-F(T)$ for fidelities $F(T)$ at the end of a ramp of timescale $T$, where $J_{1D}$ is increased from $J_{1D}=0$ to $J_{1D}=J$, with ramp function $1 - \left( e^{-\nu t} - e^{-\nu T} \right) / ( 1 - e^{-\nu T} )$, $\nu:=T/10$. We show results for $U/J=8$, for three different system sizes, $L=8$, $L=16$ and $L=32$ [symbols as for (b)].}

\label{fig:fermi_state}

\end{figure}


\emph{Summary and Outlook --} We have considered the application of a dynamical disentangling process to bosons and fermions in bilayer optical lattice systems. For realistic experimental timescales and low initial entropies, it should be possible to further suppress the entropy in a single layer by up to an order of magnitude by using this process, providing an excellent starting point for the preparation of many-body states in adiabatic processes. This could also be implemented using multiple internal states of atoms rather than spatial bilayer geometries, and the disentangling could be optimised by applying quantum control methods. On a broader level, one can ask whether such dynamical disentangling could work for a broader class of systems, opening formal questions in a quantum information context. 


\emph{Acknowledgements --} We thank Daniel Greif, Markus Greiner, Alex Ma, Marco Piani, and Jon Simon for stimulating discussions. This work was supported in part by AFOSR grant FA9550-12-1-0057, by the EOARD via AFOSR grant number FA2386-14-1-5003, and by AFOSR MURI FA9550-14-1-0035.

\bibliographystyle{apsrev}
\bibliography{bilayer}

\newpage
\clearpage

\begin{widetext}
\section*{Supplementary Material - details of dynamical disentangling}

In this section, we show that dynamical disentangling as we have described it in the main text is a controlled operation in the limit of weak coupling, meaning that the entanglement entropy between the systems is small and scales to zero with coupling, provided the original system has a gap for particle- and hole-like excitations.

We assume that the whole system consists of two subsystems, $A$ and $B$, which have a number of lattice sites $M_A$ and $M_B$ respectively, and
have associated lattice Hamiltonians $H_A$, $H_B$. We assume that these Hamiltonians conserve the number of particles in subsystems $A$ and $B$. In the absence of any coupling between $A$ and $B$, the total Hamiltonian has a separable ground state, $|\psi_0^A\psi_0^B\rangle:=|\psi_0^A\rangle\otimes |\psi_0^B\rangle$, with energy $E^A_0+E^B_0$, and also separable excited states $|\psi_{n}^A\psi_{m}^B\rangle:=|\psi_{n}^A\rangle\otimes|\psi_{m}^B\rangle$, with energies $E^A_{n}+E^B_{m}$, where $(n>0,m\geq 0) \lor (n\geq 0,m>0)$. The manifold of all excited states will in general contain a number of states that grows exponentially with $M_A$, $M_B$ and the number of particles in the system. We also assume that the total number of particles in the system is preserved by the total Hamiltonian.

Further, subsystem $A$ is assumed to have a \textit{gap} $\delta$ for particle excitations emanating into/from subsystem $B$, with well-defined particle- or hole-excitations in subsystem $A$ at energies $\geq\delta$ once these subsystems are actually coupled to each other. We will provide a technically more precise re-statement of this key assumption for our system below. We show in the following that a perturbing operator $H_c$, added to $H_A+H_B$ with prefactor $\lambda\ll \delta $, induces an entanglement $S_{A}$ ($=0$ when $\lambda=0$) between systems $A$ and $B$ that has an upper bound depending on $\lambda/\delta$, and which therefore vanishes as $\lambda/\delta\rightarrow 0$.

We assume the coupling Hamiltonian $H_c$ to be an operator sum over $I$ interface terms, 
\begin{equation}
H_c = \sum_{i=1}^I V_i^A\otimes V_i^B
\end{equation}

Here $V_i^A$, $V_i^B$ are assumed to be particle-creating or -destroying operators in subsystems $A$ and $B$ respectively, each of which changes the particle number of it's subsystem, like e.g. creation or annihilation operators in a (Bose-)Hubbard-like model. For example, in the cases treated in the main text $H_c = \sum_{x=1}^L b^\dagger_{x,1}b_{x,2} + \mbox{h.c.}$ for the Bose-Hubbard ladder, and $H_c = \sum_{x=1,\sigma=\uparrow,\downarrow}^L c^\dagger_{x,\sigma,1}c_{x,\sigma,2} + \mbox{h.c.}$ for the Hubbard ladder.  In general, $I$ will be proportional to the number of sites in subsystem $A$ directly adjacent to a site in subsystem $B$, i.e. to the physical number of interface sites $M_I$, and each $V_i^A$ and $V_i^B$ will change particle number in its subsystem quasi-locally and are always assumed to have a finite norm $\| V_i^A\|$, $\| V_i^B\|$. As a result of these two assumptions, any expectation value of the form $\langle (V^A_i)^\dagger V^A_i\rangle$ will be an intensive quantity, converging towards some finite constant in the thermodynamic limit.

If the number of particles in the total system is conserved, and $H_A$ and $H_B$ also conserve particle numbers on their own, we can always chose the manifold of all eigenvectors such that $\langle\psi_n^A|V_i^A|\psi_{n'}^A\rangle=0$  ($\langle\psi_m^B|V_i^B|\psi_{m'}^B\rangle=0$) if the number of particles in states $|\psi_n^A\rangle$ and $|\psi_{n'}^A\rangle$ ($|\psi_m^B\rangle$ and $|\psi_{m'}^B\rangle$) is the same.

To be technically explicit, what the existence of the single particle gap $\delta$ in subsystem $A$ and of well-defined particle- and hole excitations (w.r.t. excitations to/from subsystem $B$) means is two-fold. Firstly, it implies that 
\begin{equation}\label{gap_ineq}
E^A_n-E_0^A + E^B_m-E^B_0 > \delta,\quad \forall \left[ (n>0,m\geq 0) \lor (n\geq 0,m>0)\right]
\end{equation}
for all excited states that $H_c$ can couple to, i.e. for which $\langle\psi_n^A\psi_m^B|H_c|\psi_0^A\psi_0^B\rangle\neq0$
Secondly, having well-defined particle- and hole-like excitations means that the spectral functions of all $V_i^A$'s
\begin{equation}\label{defspecfunc}
A(\omega)[V_i^A] = \sum_n |\langle \psi_n^A|V_i^A|\psi_0^A\rangle|^2\delta(\omega-E^A_n+E_0^A)
\end{equation}
have appreciable support only for $f_A\times M_A$ of the exponentially many states $|\psi_n^A\rangle$ - where $f_A$ denotes a model-specific integer number, i.e. the total count of particle- and hole-like excited state modes of the system - such that the total spectral weight of these states $\sum_{k=1}^{f_A\times M_A} A(E^A_k)[V_i^A]$ converges to a finite constant if subsystem $A$ were to be taken to its thermodynamic limit. As $\int d\omega A(\omega)[V_i^A] = \langle \psi_0^A|(V_i^A)^\dagger V_i^A|\psi_0^A\rangle$ is assumed finite, the magnitude of all other matrix elements $\langle \psi_n^A|V_i^A| \psi_0^A\rangle$ for $n\notin[1,\dots,f_A\times M_A]$, denoting transitions from $|\psi_0^A\rangle$ to genuinely many-body excited states, will decay exponentially with the size of subsystem $A$. In practice, such a condition is fulfilled in a large number of gapped systems, such as a variety of MI and (pseudo-)magnetically ordered systems. In fact, the condition we have placed on~(\ref{defspecfunc}) is nothing but the re-statement of the textbook definition of a quasi-particle- or quasi-hole-like spectral function, made technically more precise to deal with the present case of a finite-sized system.

The implications for the entanglement entropy $S_{A}$ between subsystems $A$ and $B$ in the presence of the perturbing Hamiltonian $\lambda H_c$ become clear if we consider the expression for the ground state in first-order perturbation theory:
\begin{equation}\label{full_pertth1storder}
|\Psi_{\rm GS}\rangle = |\psi_0^A\psi_0^B\rangle - \lambda\sum_{\stackrel{(n>0,m\geq0) \lor (n\geq0,m>0)}{i=1,\dots,I}} |\psi_n^A \psi_m^B\rangle \frac{ \langle\psi_n^A|V_i^A| \psi_0^A\rangle  \langle\psi_m^B| V_i^B  |\psi_0^B\rangle}{E_n^A - E_0^A + E_m^B  - E_0^B } 
\end{equation}
Writing the coefficients as a matrix, $\Gamma^i_{n,m}:= \langle\psi_n^A|V_i^A| \psi_0^A\rangle  \langle\psi_m^B| V_i^B  |\psi_0^B\rangle/(E_n^A - E_0^A + E_m^B  - E_0^B)$ for every $i$, it now follows that to obtain worst-case estimates for the set of sub-leading Schmidt coefficients $\Lambda^i_n$ from a singular-value decomposition (SVD) of $\Gamma^i$, and thus $S_{A} = -\sum_{i,n}(\Lambda_n^i)^2\log((\Lambda^i_n)^2)$, we need only consider the $f_A\times M_A$ rows of this matrix that correspond to transitions from $|\psi_0^A\rangle$ to one of the $f_A\times M_A$ particle- or hole-like excited states, all other transitions being exponentially suppressed in the size of subsystem $A$. Thus, in the very worst case, each of the $I$ different matrices $\Gamma^i$ has $f_A\times M_A$ non-zero singular values, and the associated set of $f_A\times M_A$ orthogonal states for any one matrix $\Gamma^i$ will also be orthogonal to the $I-1$ sets of orthogonal states from the other $\Gamma$-matrices. If the singular values $\Lambda_n^i$ are all equal (worst case) and we take the upper boundary $1/\delta$ for $1/(E_n^A - E_0^A + E_m^B  - E_0^B)$, we arrive at $I\times f_A \times M_A$ Schmidt coefficients of size $\lambda/\delta$ and a corresponding upper bound for entanglement entropy
\begin{equation}
S_{A} \leq -\left(1-If_AM_A\frac{\lambda^2}{\delta^2}\right)\log\left(1-If_AM_A\frac{\lambda^2}{\delta^2}\right)-If_AM_A\frac{\lambda^2}{\delta^2}\log\left(\frac{\lambda^2}{\delta^2}\right).
\end{equation}

\end{widetext}

\end{document}